\documentclass[reprint,amsmath,amssymb,aps,showpacs,superscriptaddress,prb]{revtex4-1}

\usepackage{graphicx}
\usepackage{bm}
\usepackage{epsfig}
\usepackage{amssymb}
\usepackage{amsfonts}
\usepackage{braket}
\usepackage{color}
\usepackage{epstopdf}
\usepackage{hyperref}
\usepackage{float}
\restylefloat{table}
\usepackage{bibentry}
\usepackage{multirow}
\usepackage[caption=false]{subfig}
\newcommand{\ba}{\begin{eqnarray}}
\newcommand{\ea}{\end{eqnarray}}
\newcommand{\bd}{\begin{displaymath}}
\renewcommand{\v}[1]{{\bf #1}}
\newcommand{\nn}{\nonumber \\}

\graphicspath{{figures/}}% Put all figures in this directory.

\begin{document}
\title{Machine Learning Application to Two-Dimensional Dzyaloshinskii-Moriya Ferromagnets}

\author{Vinit Kumar Singh}
\email[Electronic address:$~~$]{vinitsingh911@gmail.com}
\affiliation{Department of Physics, Indian Institute of Technology, Kharagpur 721302, India}
\author{Jung Hoon Han}
\email[Electronic address:$~~$]{hanjemme@gmail.com}
\affiliation{Department of Physics, Sungkyunkwan University, Suwon 16419, Korea}
\date{\today}

\begin{abstract}
Principles of machine learning are applied to spin configurations generated by Monte Carlo method on Dzyaloshinskii-Moriya ferromagnetic models hosting the skyrmion phase in two dimensions. Successful feature predictions regarding the average spin chirality, magnetization, as well as magnetic field and temperature, were possible with the machine-learning architecture consisting of convolutional and dense neural network layers. Algorithms trained solely on the $xy$- or $z$-component of the local magnetization were as effective as the one trained on the full $xyz$ component of the input spin configuration in predicting various features. The predictive capacity of the algorithm extended beyond those configurations generated by the model used to make the training configurations, but also those generated by models plagued with disorder. A ``scaling procedure" for working with data generated at various length scales is developed, and proven to work in a manner analogous to the real-space renormalization process.
\end{abstract}
%\pacs{75.78.-n, 75.10.Hk, 75.70.Kw, 75.78.Cd}
\maketitle

\section{Introduction}

\begin{figure}[ht]
\includegraphics[scale=0.35]{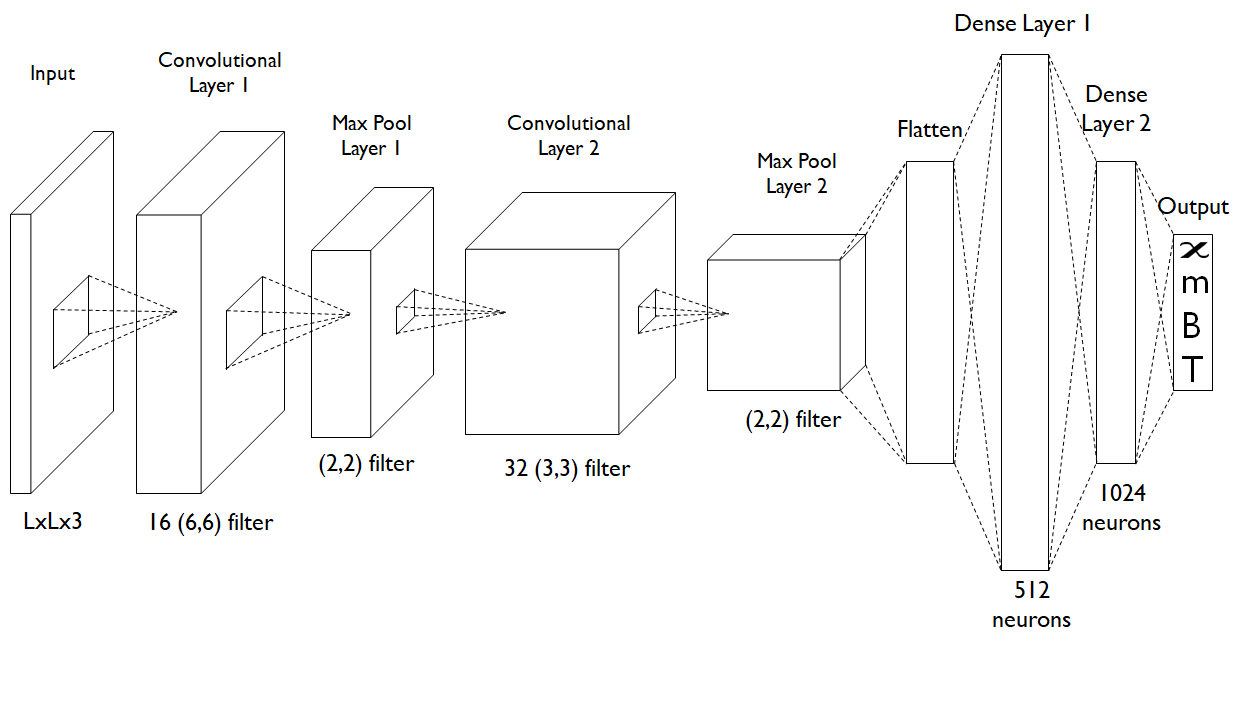}
\caption{Schematic diagram of the ML architecture used in this work. See main text for explanation.}\label{fig:1}
\end{figure}

Enormous attention has been devoted to the application of machine learning (ML) ideas to various problems of condensed matter, particularly in regard to identifying many-body phases of classical and quantum models~\cite{melko16,wang16,melko17,melko17b,melko17c,tanaka17,scalettar17,wetzel17,wetzel17b,iso18,kim18,zhai17,scalettar17,beach18,zhai18,russian18}.
Following the natural progression in the level of sophistication, models studied with the ML method have evolved from Ising~\cite{melko16,wang16,melko17,melko17b,melko17c,tanaka17,scalettar17,wetzel17,wetzel17b,iso18,kim18} to planar (XY)~\cite{zhai17,scalettar17,wetzel17b,beach18,zhai18}, and most recently to Heisenberg~\cite{russian18} spins. At first, images - either quantum wave functions or classical configurations - are generated numerically. Then such images are fed to the deep learning architecture like the one shown in Fig. \ref{fig:1}, with key physical properties such as the order parameter and topological number as the information used to ``train" the algorithm. Once the training is complete, the algorithm is capable of predicting the same physical parameters for new images, called test sets, drawn from a similar pool of circumstances but which have never been used in the training. Problems to which such supervised learning has been applied so far have found fantastic success in the machine's predictive capacity. On the other hand, it seems to be the case that the success we witnessed does not extend beyond the confirmation of knowledge already known by other means.

\section{Our model and machine-learning scheme}

\begin{figure}[h]
\includegraphics[scale=0.5]{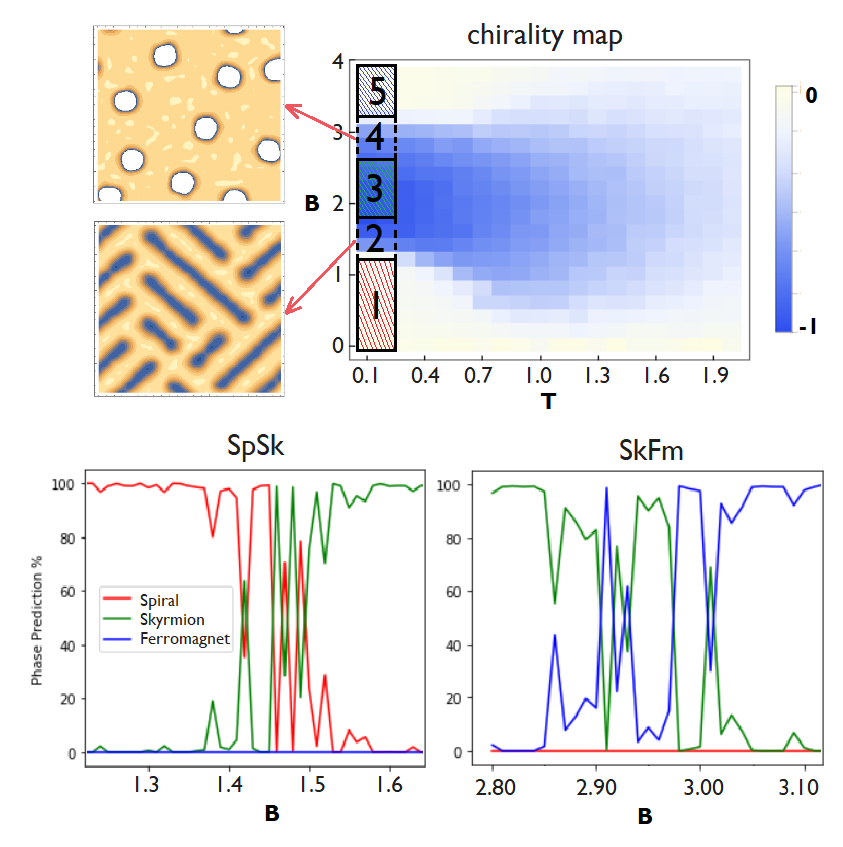}
\caption{(top) Spin chirality $\chi$ in the $(T,B)$ plane obtained by MC calculation on the HDMZ Hamiltonian (\ref{eq:HDMZ}). Color scale represents the normalized value of the chirality. Boxes 1, 3, and 5 (2 and 4) represent regions where training (testing) data were taken for label predictions. Two configurations on the left show a typical SpSk and SkFm mixed state, respectively. The $z$-component of the local magnetization is used for the plots. (bottom) Probability of phase predictions in the SpSk and SkFm phases. The numbers represent averages over the testing set in the temperature interval $T\in[0.03,0.25]$ at the same $B$ value. The irregularities are not artifacts of the small data size.}\label{fig:2}
\end{figure}

The use of ML as a supplement to the analysis of experimental data received very little attention in the condensed matter context  so far, although such has been the main thrust behind its vigorous application in the collider physics community. While it may be the over-production of data that interferes with the deduction of physical principles in the collider experiment, a typical condensed matter laboratory encounters problems of a different nature, that not all of the physical information one needs for thorough characterization of the material is available. The dearth of data is illustrated, for instance, with ultrathin materials that are amenable to various surface probes, but where thermodynamic measurement poses extreme challenge. It will be helpful, if possible, to have the ML program deduce ``missing information" on the basis of the data already at hand, without going through the painful process of measuring them. In this paper, we try to demonstrate the feasibility of such idea with a specific model, used in the study of two-dimensional spiral magnets and skyrmions:
\ba && H_{\rm HDMZ} = -J \sum_{\v r \in L^2} \v n_i \cdot (\v n_{\v r \!+\!\hat{x}} + \v n_{\v r \!+\!\hat{y}} ) \nn
 & & + D \sum_{\v r} ( \hat{y} \cdot \v n_{\v r} \! \times \! \v n_{\v r \! +\! \hat{x}} - \hat{x} \cdot \v n_{\v r} \times \v n_{\v r \! +\! \hat{y}} )  - \v B \cdot \sum_{\v r} \v n_{\v r} .  \label{eq:HDMZ} \ea
This lattice model  consisting of Heisenberg, Dzyaloshinskii - Moriya (DM) and Zeeman terms, of strengths $J$, $D$, and $B$, describes the magnetic interaction at the interface of a magnetic layer with a non-magnetic one; spins are represented as unit vector $\v n_i$ at the site $i$. Its phase diagram, by now well-established and reproduced in Fig. \ref{fig:2}, includes the skyrmion (Sk) crystal phase over some intermediate-field and low-temperature range, flanked by spiral (Sp) phase at low field and ferromagnetic (Fm) phase at high field~\cite{nagaosa-review,skyrmion-book,jiang-review,fert-review,han-book}. The spiral state has the period $\lambda$ fixed by the ratio $D/J$~\cite{han09,han-book} according to $\sqrt{2}\tan (2\pi/\lambda)  = D/J$, which also serves as the diameter of the skyrmion. The lattice spacing $a$ on the $L\times L$ square lattice is taken to unity in the model calculation, while it is a few \AA~in actual materials.  The typical spiral and skyrmion size is several tens and hundreds of lattice constants due to the small ratio $D/J \ll 1$, but it is customary in model calculations to assume $D/J$ corresponding to a few lattice constants. One justifies this on the basis that, although physical systems with their long-period structures are best described by a continuum free energy functional, in numerical studies one discretizes the continuum model and put it into the form (\ref{eq:HDMZ}) with the lattice spacing $a$ having no direct relation to the underlying physical lattice constant (see Ref. \onlinecite{han09,han-book} for details of the discretization procedure). The lattice model parameters $J, D, B$ in (\ref{eq:HDMZ}) are renormalized from their original meaning in the continuum theory and now have the unit of energy, as does the temperature $T$ to be introduced in later Monte Carlo simulation. Conversion to physical temperature and magnetic field scales can be done with the aid of Boltzmann's constant and the Bohr magneton. Skyrmions are characterized by the topological charge, equal to the integral (or sum, if on a discrete lattice) of the spin chirality. The importance of topologically protected skyrmions as information carriers has received enormous attention recently. Readers interested in further background on skyrmions can follow recent publication of books and reviews~\cite{nagaosa-review,skyrmion-book,jiang-review,fert-review,han-book}. Surprisingly, very little attention has been given to the utility of the ML scheme toward the analysis of skyrmion experiments (Ref. \cite{russian18} focused on the phase identification problem instead).

Figure \ref{fig:1} shows the supervised ML architecture used in this work. The input size is $(L,L,n)$ for $L\times L$ lattice.  The training data was generated by running Monte Carlo (MC) simulation on the model Hamiltonian (\ref{eq:HDMZ}) with $D/J=\sqrt{6}$, corresponding to the spiral period $\lambda=6$. This represents a convenient choice of length scale, not necessarily corresponding to actual period of the spiral in the experiment. Later we will show how the data generated at other length scales can be coarse-grained and transformed to a data set with  ``effective" lattice spacing equal to 6. We performed the ML training in three different ways using the $z$-component, $xy$-components, and full $xyz$-components of the magnetization vector $\v n_i$ generated by MC simulation. Roughly 200 training configurations are generated for each point of the $(B, T)$ grid covering the entire phase diagram shown in Fig. \ref{fig:2}: $0 \le B \le 4$ and $0.03 \le T \le 2$. For each $B$, the temperature interval was divided into 40 steps using adaptive scheduling, {\it i.e.} exponentially decaying step size with a decay rate of 0.1, giving a total of 20 steps. With 17 uniformly spaced magnetic field values, we drew 100 MC configurations at each $(B,T)$, for a total of $20 \times 100\times 17 = 34,000$ training sets. Training with much larger training set of 330,000 configurations did not change the final results. Both the fineness of the grid spacing and the number of training configurations were such that no further improvement in the performance was possible.
Testing configurations are picked from the same model, but with a separate MC run to generate them.

To ensure that the natural periodicity of the model (\ref{eq:HDMZ}) is faithfully understood by the machine, an initial convolutional neural network (CNN) layer with 16 filters, each of size 6$\times$6, was used. It was followed by the Max Pool layer of 2$\times$2 filter size, then by a second CNN layer with 32 (3$\times$3) filters.  The 3$\times$3 was chosen out of trial-and-error for the best results. Batch Normalization and Dropout Regularization accompanied both CNN layers. After applying the second Max Pooling to reduce the size, the data was fed through two Dense Neural Network (DNN) layers containing 512, 1024 neurons respectively, which then led to the output layer. Batch normalization and Dropout Regularization are applied to outputs from each DNN layer. Leaky ReLu was employed as the activation function with $\alpha=0.1$, except for the output layer where a sigmoid function was used. Adam optimizer and Learning Rate Scheduler were applied to enhance the training speed.  The training input data was arranged in terms of the local unit vector $\v n_i$, not in terms of the two angles which characterize it, due to the poorer performance in the latter case. The architecture consisting purely of the DNN layer as in Ref. \onlinecite{russian18} generally did not work as well as the one involving also the CNN filter layers. Further minute changes in the architecture had little impact on the overall quality of final results. Exhaustive discussion of the CNN, DNN, and other nomenclature can be found in several recent books~\cite{bishop,goodfellow} and on online courses~\cite{ng}.

Although the phase identification problem is not the main issue in this work as other reports already exist~\cite{russian18}, we do note an interesting aspect of the model (\ref{eq:HDMZ}) which is absent in most models studied in recent years and not amply addressed in the previous work~\cite{russian18}. Due to the substantial co-existence region both for the mixture of spiral and skyrmion (SpSk), and of skyrmion and ferromagnetic (SkFm) phases, we can categorize the overall low-temperature phase diagram in terms of five states (Sp, SpSk, Sk, SkFm, Fm), indicated as boxes numbered 1 through 5 in Fig. \ref{fig:2}. To see how the ML would fare against such a conspicuous mixed phase, we first did the ML training for MC configurations drawn from boxed regions 1, 3, 5 (three pure states). One-hot encoded labels were used to represent different phases: (1,0,0) for Sp, (0,1,0) for Sk, and (0,0,1) for Fm. Afterwards, configurations drawn from boxes 2, 4 (two mixed states) were fed to the program, demanding that it decides which of the three phases they belong to. Binary cross entropy was used as the loss function. The answers given for each test configuration by the machine were averaged over and shown as probabilities for Sp, Sk, and Fm phases  in Fig. \ref{fig:2}. Despite the fact that each data point in the figure represents an average over $2,000$ test configurations and that extremely fine steps in magnetic field $\Delta B= 0.01$ was used, the final results are far from being smooth. Removing the CNN filters did not smooth the outcome either. In contrast, a smooth variation in the probability from 1 (ordered phase) to 0 (disordered phase) was found in models with a second-order phase transition \cite{wang16,melko17,tanaka17,scalettar17,wetzel17,kim18,zhai17,scalettar17,beach18}.
The ``failure" of the ML algorithm in recognizing and characterizing mixed phases around the first-order phase transition was not appreciated in earlier investigation of the same model~\cite{russian18}.

\section{feature prediction scheme}

\begin{figure}[h]
\includegraphics[scale=0.35]{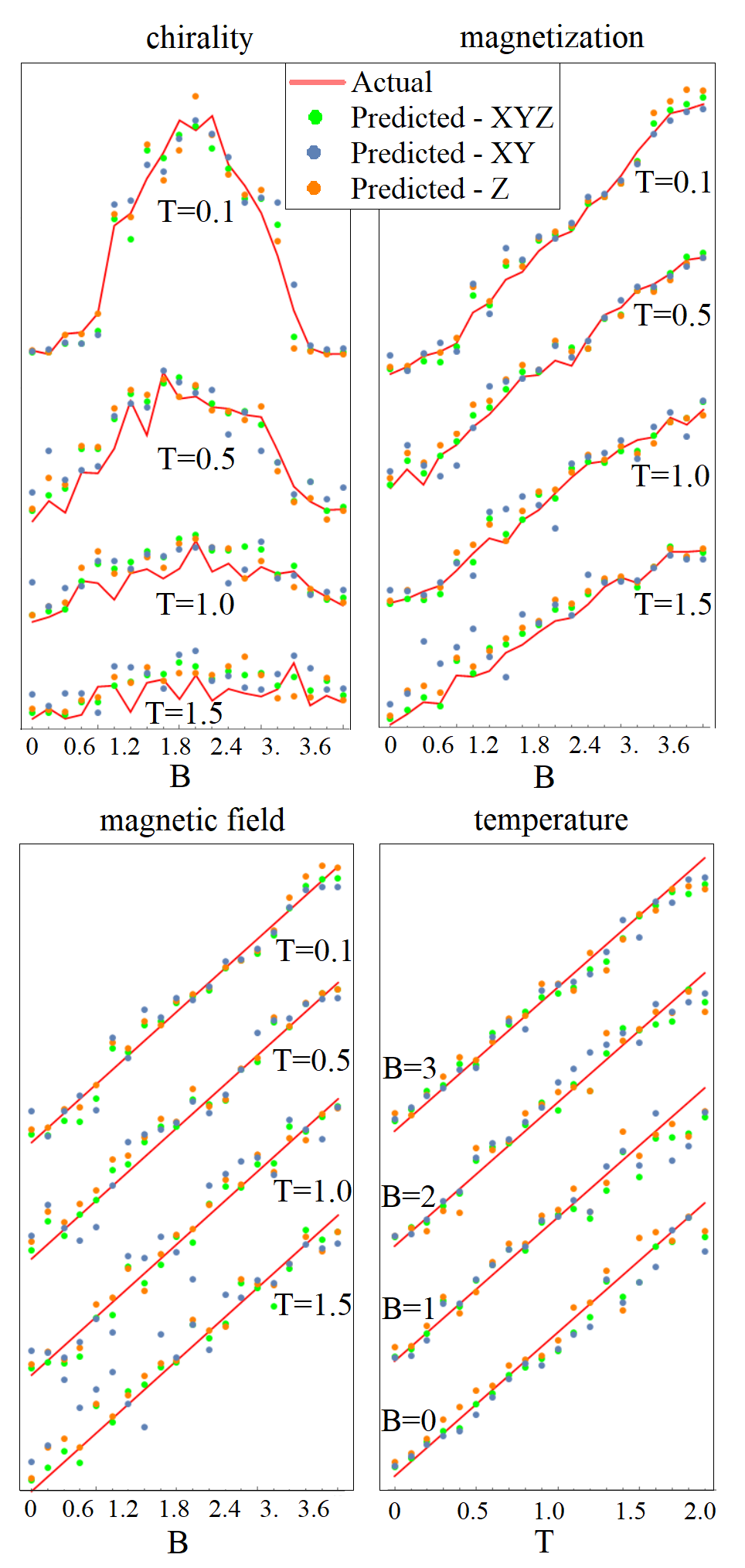}
\caption{Machine-predicted values of $(\chi, m, B, T)$ compared to their actual values in red. Predicted values are obtained from algorithms trained on the full $xyz$, $xy$-only, and $z$-only components of the local magnetization. Different curves are offset for clarity.}\label{fig:3}
\end{figure}

The two main order parameters of the phase diagram in the model (\ref{eq:HDMZ}) are the average spin chirality and the magnetization; the spin chirality becomes prominent in the skyrmion phase, the magnetization in the ferromagnetic phase, and the spiral phase supports neither. A discrete version of the spin chirality is the solid angle subtended by three adjacent spins, given by~\cite{berg81,zang16,han-book}
\ba \tan \left( {\chi_{123} \over 2} \right)  = {\v n_1 \cdot \v n_2 \times \v n_3 \over 1 + \v n_1 \cdot \v n_2 + \v n_2 \cdot \v n_3 + \v n_3 \cdot \v n_1 }  . \label{eq:discrete-chi}\ea
In the smoothly-varying limit ($\v n_i \cdot \v n_j \approx 1$) one recovers the familiar expression $\chi_{ijk} = \v n_i \cdot \v n_j \times \v n_k /2$, which upon taking $i=\v r$, $j = \v r + a \hat{x}$ and $k = \v r + a\hat{y}$ and $a\rightarrow 0$ gives $\chi_{123} = a^2 \v n \cdot (\partial_x \v n \times \partial_y \v n)/2$. The spatial averages $\sum_{i} ( \chi_{\v r, \v r\!+\!\hat{x}, \v r \!+\!\hat{y}} + \chi_{\v r , \v r \!-\!\hat{x}, \v r \!-\! \hat{y}} ) /L^2 =\chi$ and $m =\sum_{\v r} n_{\v r}^z /L^2$  define the spin chirality and the magnetization, respectively. We propose to train the algorithm on the {\it features} of the data such as $\chi$ and $m$ (mechanical quantities), as well as the magnetic field $B$ and temperature $T$ (thermodynamic quantities). Here, the mean-squared error function was used as the loss function. Training configurations were drawn from the entire phase diagram in Fig. \ref{fig:2}. After training, 40,000 configurations were freshly generated to compare the machine-predicted $(\chi,m, B, T)$ against their actual values. Figure \ref{fig:3} gives the comparison of the original and machine-predicted $(\chi, m, B, T)$. Good agreements are found on all four quantities. Estimation of the error is given in the Appendix. We obtained very similar quality of errors from both definitions of the spin chirality (\ref{eq:discrete-chi}) or its smooth form $\chi_{ijk} = \v n_i \cdot \v n_j \times \v n_k /2$ (Note an alternative definition of the spin chirality was adopted in Ref.~\onlinecite{pujol15}). Results shown in the figures are based on the smooth form, which is slightly easier to compute numerically. In short, the algorithm has successfully figured out a recipe to extract $(\chi, m, B, T)$ values implicit in the spin configuration. Most remarkably, equally good predictions were possible for all three types of training data sets -  $xyz$, $xy$, and $z$ - meaning that the test image consisting of only the $z$-component of local magnetization was sufficient to predict the correct spin chirality $\chi$, which ordinarily requires the full knowledge of spin components for computation.

The successful prediction of spin chirality based on $xy$- and $z$-only inputs, or of magnetization based on $xy$-input alone, holds interesting potential for application of ML program in the actual experiment.
Consider two prominent imaging techniques currently under use to study skyrmion matter: Lorentz transmission electron microscopy (LTEM) \cite{tokura10} and magnetic force microscopy (MFM) \cite{pana17}. Each of them specializes in imaging the in-plane $(n^x_{\v r} , n^y_{\v r} )$ and perpendicular ($n^z_{\v r}$) components of the local magnetization. The data provided by LTEM and MFM is thus $xy$- and $z$-type, respectively.  Given the raw data, it is impossible to determine the spin chirality directly, nor to deduce the average magnetization from the LTEM data. With the ML program we developed, it appears within reach to compute both $\chi$ and $m$ with existing, partial experimental inputs. Achieving this, however, would require the establishment of a careful protocol by which the raw experimental data is converted to the machine-ready data set.

\begin{figure}[t]
\includegraphics[scale=0.36]{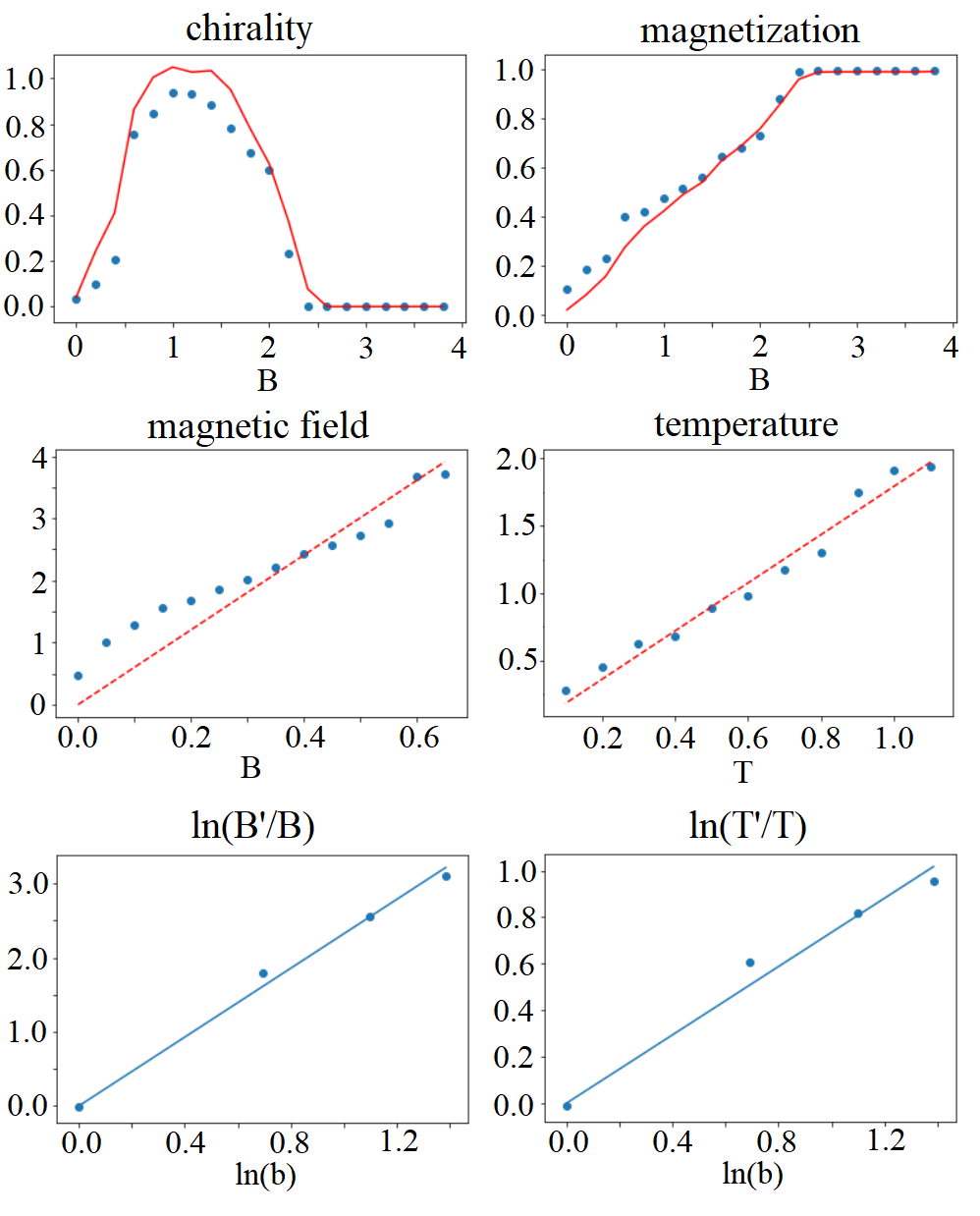}
\caption{Top row: Machine-predicted values of $(\chi', m' )$ after coarse-graining and rescaling the original data with $b=2$. Red curve is from applying $\chi$ and $m$ formulas to pre-scaling data; blue dots are machine predictions based on renormalized input of $\v n'_i$. Middle row: Machine-predicted values $(B', T')$ vs. $(B, T)$ for $b=2$. $B'/B$ and $T'/T$ are approximately constant. Bottom row: Inference of scaling exponents from $\ln (B'/B )$ (and $\ln (T' /T)$ vs. $\ln b$ plots. Least-square fit gives slopes 2.32 and 0.73, respectively. } \label{fig:5}
\end{figure}

An obvious obstacle seems to be the vastly different length scales between experimentally-available and theoretically-generated data sets. For one, the typical size of actual skyrmion or spiral, measured in terms of physical unit-cell spacings, is 1-2 orders of magnitude greater than the period-6 structure used in our training. One way to bridge the gap in the length scales is to use coarse-graining procedure on the experimental data to match the renormalized length scale with the one used in the training. This way, the LTEM/MFM data would become compatible with the existing algorithm. Access to raw experimental data is currently not available to the authors, but one can opt for the following  proxy: in the previous ML tryouts, both the training and the test sets were generated with the same period ($D/J=\sqrt{6}$); this time, we adopt several $D/J$ values compatible with the spiral periods $\lambda=12, 18, 24$, and generate MC configurations on $L=48, 72, 96$ lattices, respectively. The MC configurations are subsequently coarse-grained according to the simple scheme, $\v n'_{\v r} = \sum'_{\v r} \v n_{\v r} / | \sum'_i \v n_{\v r}  |$, where the sum $\sum'_{\v r}$ is over the $b\times b$ block of original spins, to produce the renormalized spin configuration $\v n'_{\v r}$ on $24\times24$ lattice with the common renormalized period 6 by using $b=2,3,4$ for $\lambda=12, 18, 24$, respectively. We pose a scaling hypothesis akin to the one in real-space renormalization; one anticipates the predictions $(\chi',  m',  B' , T')$ for the renormalized spins $\{ \v n'_i \}$ to be related to $(\chi, m, B, T)$ of the original spin configuration $\v n_i$ through some scaling relation, e.g. $\chi'/\chi = b^{\#}, B'/B = b^{\#}$ with appropriate exponents $\#$. As shown in Fig. \ref{fig:5} we find $\chi' \approx \chi$ and $m' \approx m$ independent of $b$, but the ratio  $B'/B$  and $T'/T$ obey an approximate scaling relation $B'/B \approx b^{2.32}$ and $T'/T \approx b^{0.73}$. The linear fit on a log-log scale is fairly good (errors are estimated in the Appendix), which brings us to suspect that scaling should work even at other, non-integer values of $b$. It means that even with just one period $\lambda$ used in the training, the ML program is capable of producing reliable predictions for MC configurations - and even experimental data - at other, arbitrary  $\lambda$.

Experimental situation is still more complicated than what the simplified model (\ref{eq:HDMZ}) captures. A most obvious complication is the disorder effect, coming from inhomogeneities in the interaction parameters of the Hamiltonian or local anisotropy terms we ignored so far. The predictive capacity of ML would be much reduced if the it works well only on test  configurations generated by the exact same model from which training set was generated in the first place. Luckily, the numerical experiment presented below shows that the ML program trained on the pristine model continues to give reliable predictions for test configurations obtained from models with disorder. Replace the dirty model configurations by the actual experimental images (which always come from imperfect samples), we come to conclude that ML could give reliable predictions from messy experimental data as well.

\begin{figure}[t]
\includegraphics[scale=0.36]{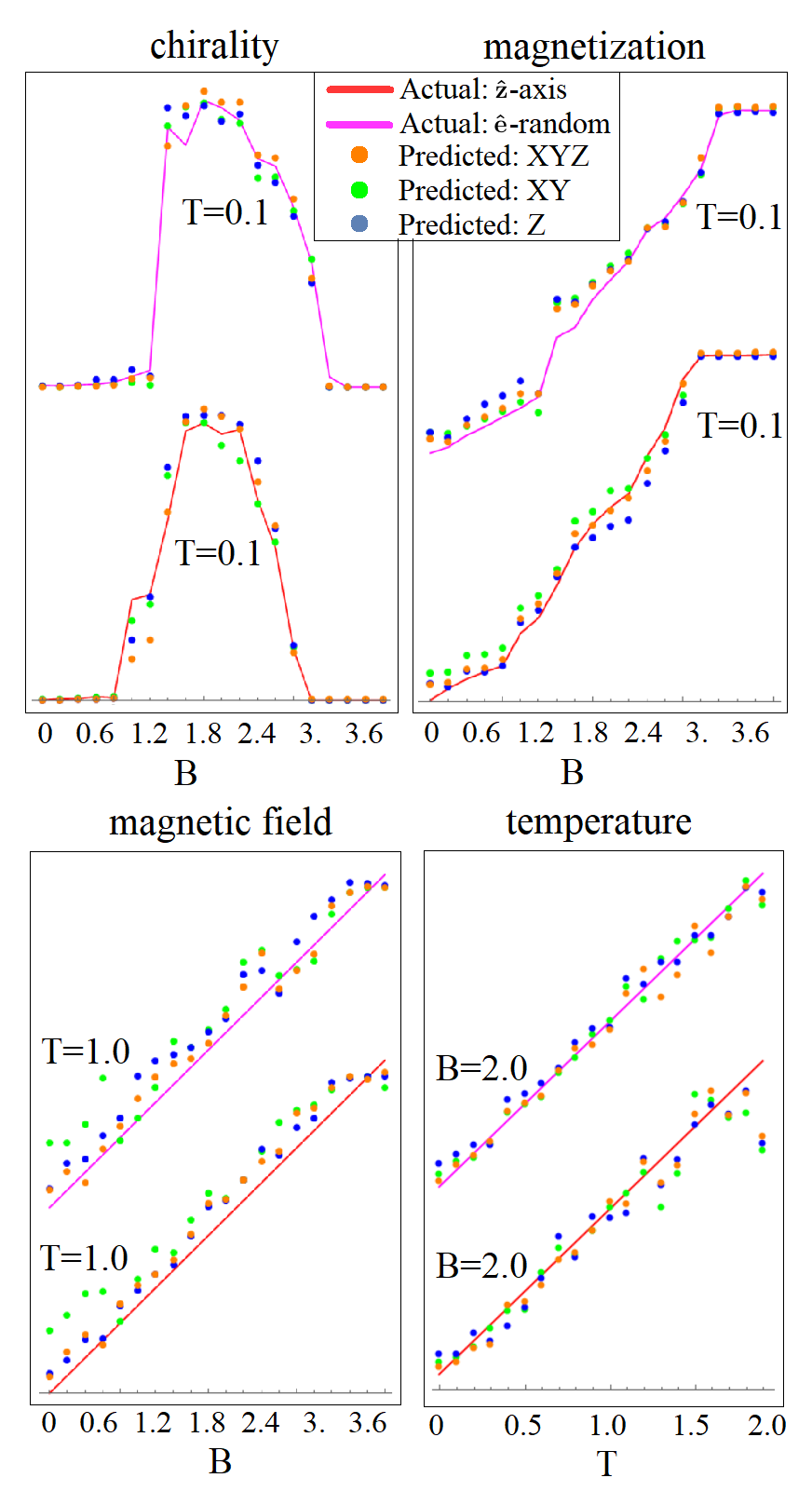}
\caption{ML prediction of $(\chi, m, B, T)$ for MC configurations generated by $H_{\rm HDMZ}+ H_1$, and $H_{\rm HDMZ} + H_2$, with $(K,p)=(1, 0.5)$ for both models.  All three input data types ($z$, $xy$, $xyz$) give equally good predictions. Error estimation is given in the Appendix. } \label{fig:4}
\end{figure}

We consider two kinds of disorder terms to add to the HDMZ Hamiltonian (\ref{eq:HDMZ}):
\ba H_{1} = -K \sum_{\v r \in {\rm ran}} (\hat{z} \cdot \v n_{\v r} )^2 , ~~ H_2 = - K \sum_{\v r \in {\rm ran}} (\hat{e}_{\v r} \cdot \v n_{\v r} )^2 .  \ea
In $H_1$, easy $z$-axis magnetic anisotropy of strength $K$ is added at the random sites occupying a fraction $p$ of the whole lattice. In $H_2$, the anisotropy orientation $\hat{e}_i$ is also random. A new batch of test configurations has been generated by doing MC on either  $H_{\rm HDMZ} + H_1$ or $H_{\rm HDMZ} + H_2$. We do not, however, generate a new ML algorithm trained on these new configurations. The newly-generated MC configurations are simply fed to the existing program, trained on the disorder-free model, and predictions for $(\chi, m, B, T)$ are demanded. As before, three different input types ($z$, $xy$, and $xyz$) have been used on three different ML programs. Figure \ref{fig:4} shows the outcome of such investigation. In essence, actual $(\chi, m, B, T)$ values of the test configurations with disorder are faithfully reproduced across all three data and ML types. Our numerical experiment suggests that the predictive power of ML remains universal across a range of disorder potentials. Overall errors in the predicted values of $(\chi, m, B, T)$ against the actual ones are tabulated in the Appendix.

\section{How does the machine learn?}

The ML code we have developed has done well, it seems, as far as predicting certain physical quantities goes. On the other hand, we do not understand clearly the inner workings of this ``black box" which makes such excellent prediction possible. One could say that in a way the machine has ``understood" the concepts of average magnetization, spin chirality, and so on, during the process of training, and has learned to apply such concepts to new circumstances. It is even remarkable that the spin chirality prediction was no less successful for inputs consisting of only partial spin components, {i.e.} either $xy$- or $z$-components only. We ask, in this section, if such learning process of the machine program can be uncovered by looking closely at the intermediate processes in the overall ML procedure outlined in Fig. \ref{fig:1}.

We start with the CNN layer. The CNN scheme has shown excellent performance in many computer vision and machine learning problems. From there we got the idea that it could perform equally well on data obtained from physical experiments like MFM and LTEM. The input to a CNN is either an experimentally obtained or simulation-generated image that can be expressed, in our case, as a data of spins with three components $(n^x_{\v r} , n^y_{\v r} , n^z_{\v r})$ over $\v r \in L\times L$ lattice. The input data then sequentially goes through a series of processing, as depicted in Fig. \ref{fig:1}, to yield a physical answer. Each processing is done via an operator which are usually matrices.

The initial values of these matrices are random, hence after applying these processing steps on input image one would initially find a wrong result. Then the machine compares the machine-predicted result with the actual result called the ``labels", which in our case corresponds to chirality or magnetization. The algorithm then consistently updates the operation matrix elements until it best predicts the final results by using various techniques of error minimization, {\it e.g.} Forward Propagation, Backward Propagation and Stochastic Gradient Descent. When the machine starts predicting correct results after adjustments of the matrix elements, the network is said to have been trained.

Given a one-dimensional array of input data, denoted $f(m)$ where $m$ are integer indices, the convolution takes place by multiplying it with a filter function $g(n)$ in the manner called the {\it convolution}:
\ba (f\circ g)[n] = \sum_{m} g(n-m) f(m)  . \nonumber \ea
Higher-dimensional version of the convolution is possible, which is how the CNN filter works mathematically. For three-dimensional
input data $f(i,j,k)$ the convolution operation in the CNN works as
\ba (f\circ g)[i,j,k] \!=\! \sum_{l,m,n=-l_f }^{l_f} g(i\!-\!l,j\!-\!m,k\!-\!n) f[ l,m,n ] , \nonumber \ea
where $l_f$ is the filter size. Elements of the filter function $g$ are determined self-consistently during iterative minimization of the error. Through such convolution operation the input image, or the input set of numbers, become transformed into a new set of numbers, often of dimensionality different from the original input. There is more than one filter function $g$ employed in the CNN procedure; in the case of 16 CNN filters, 16 different kinds of filter functions $g$ are developed in order to capture the various aspects of the input image.

The notion of a filter can be demonstrated with simple examples. The process of calculating the local magnetization $n_{\v r}^z$ is equivalent to applying a filter function $g(i,j,k) = \delta_{i,0}\delta_{j,0}\delta_{k,3}$. Here $\v r = (i,j)$ refers to the spatial position of the spin, and $k=1,2,3$ refers to the $x,y,z$ component of the input spin. By using such a filter one selects $(f\circ g)[i,j,k] = f(i,j,3)$, namely the $z$-component of the local magnetization at the site $\v r$. A more complicated example is the local spin chirality, $\chi_{\v r } = \v n_{\v r} \cdot (\v n_{\v r +\hat{x}} \times \v n_{\v r+\hat{y}})$. The filter function for the spin chirality can be defined as $g_{\v r_1, \v r_2, \v r_3}^{\alpha\beta\gamma} = \delta_{\v r_1 , 0} \delta_{\v r_2 , \hat{x}} \delta_{\v r_3, \hat{y}} \varepsilon^{\alpha\beta\gamma}$, thus its application on the spin configuration $n_{\v r}^\alpha n_{\v r_2}^\beta n_{\v r_3}^\gamma$ should yield the spin chirality. The CNN architecture learns to create such filter function without the {\it a priori} knowledge of its mathematical form, through repeated test of the predicted values of spin chirality and magnetization against the actual values.

With this elementary concept of CNN layer in mind, we attempt to demonstrate the actual inner workings of the CNN and the subsequent DNN with the help of examples as shown in Fig. \ref{fig:6}. It illustrates the full ``journey" of an input spin data through the neural network and how it is finally transformed to yield the four basic physical parameters, {\it i.e.} chirality, magnetization, magnetic field, and temperature.  Recall that we used input images from three distinct phases of the model: spiral, skyrmion, and ferromagnet. Furthermore, we used three different input types, consisting of full $xyz$, $xy$-only, and $z$-only components of spin, only to find very similar physical parameters at the end of their respective numerical journey. Each column in the image of Fig. \ref{fig:6} corresponds to one such specific input. Different rows in Fig. \ref{fig:6} in turn correspond to successive CNN and DNN layers. Detailed explanations of the flow from input to output can be found in the figure caption as well. 

Overall, there is certain visual analogy between the input image and some of the CNN images in the first layer, while the second-layer CNN images have become more abstract. Once the CNN images are converted to a string of numbers in the DNN layer, the pictorial aspect of the input image is completely gone; instead all the relevant information that have been extracted from the input image are now expressed in the string of numbers. One may draw certain analogy to a musical recording buried in the string of 0's and 1's on a physical compact disc. The string of bits does not look like music at first sight, but our daily experience amply demonstrates their equivalence.  Once the image file has been condensed into a string of 1024 numbers in the DNN2 layer, physical outcome such as magnetization will come out by taking certain combinations of them. The whole procedure is well-defined numerically, but unlike the formulas in physical sciences, it is hard to attach certain intuition to the mathematical procedure with which the CNN and DNN layers process the physical information.

\section{Conclusion}
In this paper, we have demonstrated the versatility of the machine learning program in predicting physical quantities of interest in models hosting the skyrmion phase. The average magnetization and the spin chirality are the two key quantities characterizing the phases of the skyrmion matter, and both proved to be trained with high accuracy by the standard machine-learning procedure. The robustness of the predictive power was further tested in various ways. The input data consisting of only partial information of the spin orientation proved to be adequate in successfully predicting physical quantities. Predictions made on a data set generated by the Hamiltonian which was perturbed by impurity effects were also quite accurate. Even the concept of scaling was proven to work in the machine learning prediction, which can be a powerful hint that the machine learning program can be used in conjunction with the experimental probe to determine the magnetic structure with more precision. While the ultimate utility of the protocol we propose  has not yet happened in the real-world application, we argue that the theoretical support given in this paper clearly suggests the such experiment-ML collaboration will be rewarding.

\acknowledgments
This work was supported by Samsung Science and Technology Foundation under Project Number SSTF-BA1701-07.
J. H. H. thanks Manhyung Han for instructive conversations on machine learning.

\appendix
\section{Summary of error analysis}

\begin{table}[htb]
\begin{tabular}{ | ccc || ccc  cc  cc  cc |}
\hline
 &  XYZ-type & & & $\Delta\chi$ &  & $\Delta m$ &  & $\Delta B$ & & $\Delta T$ & \\ \hline
 &  $H_{\rm HDMZ}$ & & & 5.82 &  & 3.79 &  & 4.91 & & 5.32 & \\ \hline
 & $H_{\rm HDMZ}+ H_1$  &  & & 5.83 & & 3.85 & & 5.49 & &  5.62 & \\ \hline
 & $H_{\rm HDMZ}+ H_2$ & & & 6.01 & & 3.77 & & 7.22 & & 6.75 & \\ \hline
 & $H_{\rm HDMZ} ~ (b=2)$   & & & 7.05 & & 4.15 & & 10.1 & & 4.66 & \\ \hline
 & $H_{\rm HDMZ} ~ (b=3)$ & & & 6.46 & & 3.69 & & 11.7 & & 4.93 & \\ \hline
 & $H_{\rm HDMZ} ~ (b=4)$ & & & 6.61 & & 4.07 & & 12.2 & & 5.89 & \\ \hline \hline
 &  XY-type & & & $\Delta\chi$ &  & $\Delta m$ &  & $\Delta B$ & & $\Delta T$ & \\ \hline
 &  $H_{\rm HDMZ}$ & & & 7.15 &  & 5.4 &  & 7.28 & & 5.23 & \\ \hline
 & $H_{\rm HDMZ}+ H_1$  &  & & 7.52 & & 6.2 & & 8.5 & &  5.42 & \\ \hline
 & $H_{\rm HDMZ}+ H_2$ & & & 8.25 & & 7.76 & & 11.8 & & 6.37 & \\ \hline \hline
 &  Z-type & & & $\Delta\chi$ &  & $\Delta m$ &  & $\Delta B$ & & $\Delta T$ & \\ \hline
 &  $H_{\rm HDMZ}$ & & & 5.98 &  & 3.28 &  & 5.14 & & 6.33 & \\ \hline
 & $H_{\rm HDMZ}+ H_1$  &  & & 6.09 & & 3.2 & & 5.56 & &  6.48 & \\ \hline
 & $H_{\rm HDMZ}+ H_2$ & & & 5.65 & & 3 & & 7.2 & & 6.66 & \\ \hline
\end{tabular}\label{table:1}
\caption{Averaged variance between predicted and actual values of $(\chi, m, B, T)$.}
\end{table}

Listed in Table 1 are the errors in the machine-predicted values of $(\chi, m, B, T)$. The error estimation is done by the formula
\ba \Delta X = \sqrt{ { \sum_i (X_{{\rm predicted}, i} - X_{{\rm actual}, i} )^2 \over N} }. \ea
Here $X=\chi, m, B, T$ and $1\le i \le N$ ranges over all the test configurations. Input data types are classified as $xyz$, $xy$, and $z$, according to all three components, only $xy$-component, and only $z$-component of the local magnetization vector $\v n_i$ being used for training and testing. The pure case $H_{\rm HDMZ}$ refers to the choice $D/J=\sqrt{6}$ corresponding to the spiral period $\lambda=6$. The two disordered Hamiltonians we considered in the main text are shown in the rows with $H_{\rm HDMZ}+ H_1$ and $H_{\rm HDMZ}+ H_2$. The sample size is $N= 20\times 20 \times 100$.

For $b=2,3,4$, only the pure Hamiltonian $H_{\rm HDMZ}$ was used with $D/J$ values corresponding to $\lambda=12,18,24$, respectively. The resulting raw data is compressed according to the block-spin rule (mentioned in the text) before being subject to machine prediction. The predicted values of $\chi, m, b, T$ are then compared to $\chi', m', B', T'$, which is related to the raw value through the scaling relation $\chi'/\chi = b^\#$. The exponents used are 0, 0, 2.32, and 0.73, respectively. For example, the variance $\Delta B$ in the case of $b=2$ is obtained from
\ba \Delta B = \sqrt{ { \sum_i (B_{{\rm predicted}, i} - B_{{\rm actual}, i} 2^{2.32} )^2 \over N} } \ea
where $B_{{\rm actual}, i}$ is the magnetic field used in the generation of the $\lambda=12$ Monte Carlo configuration. The sample size was $N=14\times 11 ~ (b=2)$, $N=14\times 9 ~ (b=3)$, and $N=15\times 7 ~ (b=4)$.

\clearpage

\begin{widetext}
\begin{figure}[t]
\includegraphics[scale=0.8]{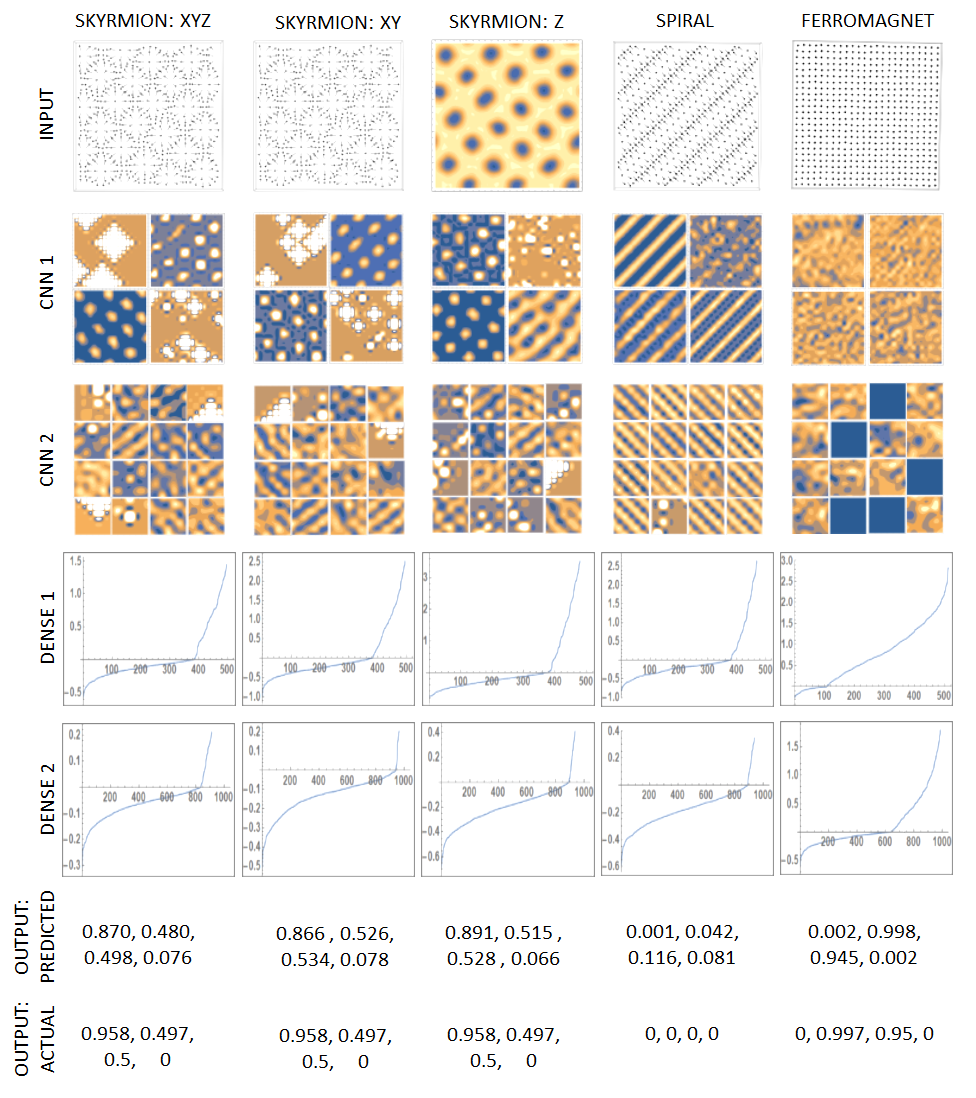}
\caption{(first row) Input layer: It shows a graphical representation of the input spin configuration on $L\times L$ lattice ($L=24$ in this figure). The first three columns show images of input skyrmion configuration consisting of $xyz$, $xy$-only, and $z$-only components of the spin. The color scheme in the $z$-only case represents $n^z_i$ as colors. Bright means spin up, and dark means spin down. For other input types, explicit spin components are plotted. (second row) CNN1 layer: The input layer is operated on with a convolution matrix with sixteen CNN filters, each of size $6\times 6$, to yield images of size $(L-5)\times (L-5)$. Four out of sixteen such images are displayed here for each input spin configuration. (third row) CNN2 layer: Images from the CNN1 layer are operated upon by thirty-two CNN filters, each of $3\times3$ size, to yield images of size $[(L-5)/2] -2 \times [(L-5)/2] -2 = 7 \times 7$. Sixteen out of thirty-two such images are displayed. (fourth row) Dense layer 1: Having passed through the CNN layers, the image (more precisely, the numbers which define the image) is flattened into a one-dimensional array of numbers, and then transformed into an array of length 512 through multiplication of weight matrix. The 512 numbers in an array are plotted in ascending order. The overall vertical scale is arbitrary. (fifth row) Dense layer 2: The 512-length array in dense layer 1 is expanded into another dense layer with 1024 numbers shown in the ascending order. Finally, these 1024 numbers are transformed to yield four numbers which are of physical importance - chirality, magnetization, magnetic field, and temperature - shown in the sixth row and compared against actual values in the seventh row.} \label{fig:6}
\end{figure}
\end{widetext}

%\bibliographystyle{apsrev}
%\bibliography{reference}

\end{document}